\documentclass[%
preprint,
superscriptaddress,
 amsmath,amssymb,
 aps,
floatfix,
]{revtex4-2}

\usepackage[normalem]{ulem}
\usepackage[dvipsnames]{xcolor}
\usepackage{graphicx}
\usepackage{dcolumn}
\usepackage{bm}
\usepackage{amsmath}
\usepackage{mathrsfs}

\begin{document}

\title{Breaking Space Inversion-Symmetry to Obtain Asymmetric Spin-Wave Excitation in Systems with Nonuniform Magnetic Exchange}



\author{Rair Mac\^edo*}
\affiliation{James Watt School of Engineering, Electronics \& Nanoscale Engineering Division, University of Glasgow, Glasgow G12 8QQ, United Kingdom}
\email{Rair.Macedo@glasgow.ac.uk}

\author{Arjun S. Kudinoor}
\affiliation{Department of Physics, Columbia University, New York, New York 10027, United States}

\author{Karen L. Livesey}
\affiliation{School of Mathematical and Physical Sciences, The University of Newcastle, Callaghan NSW 2308, Australia
}

\author{Robert E. Camley}
\affiliation{Center for Magnetism and Magnetic Materials, Department of Physics and Energy Science, University of Colorado Colorado Springs, Colorado Springs, Colorado 80918, USA}

\date{\today}

\begin{abstract}

We report on the consequences of non-uniform exchange in magnetic systems. 
The quantum mechanical exchange interaction between spins is responsible for the phenomenon of magnetic order, and is generally considered to be uniform across bulk magnetic systems.
Partly inspired by the Dzyaloshinskii-Moriya interaction{\textemdash}also known as antisymmetric exchange{\textemdash}we use a linearly varying exchange interaction along a magnetic strip as a route to spatial inversion symmetry-breaking. 
We find that, in addition to asymmetric modes and localization, spatially-varying exchange can be used to design nonreciprocal magnetic signal excitation at frequencies that are tunable. 
Moreover, our work predicts nonreciprocity to occur across a vast range of frequencies up to hundreds of GHz.
Such spin wave engineering is a key area of ongoing research in the fields of magnonics and spintronics, which are expected to enable the next generation of communication technology. Analogous nonreciprocity is expected to occur in other wave systems with gradient properties.
\end{abstract}


\maketitle



\section{\label{sec:into}Introduction}

Much attention has been paid to the control and manipulation of spin waves in magnetic materials \cite{chumak15}.
This is because controlling spin waves is the basis for achieving, for example, multifunctional magnonic circuits to perform a myriad of tasks and significantly advance the capabilities of computation \cite{mahmoud20}, data processing, and data storage technologies \cite{chumak17}.
In addition to new functionality, spin-wave-based devices may also have reduced energy requirements \cite{haldar16}.
Artificial magnetic materials such as exchange springs and magnonic crystals have been designed to have spin waves with required properties.

Here, we investigate a magnetic system wherein spatial inversion symmetry is broken by making the exchange interaction between spins nonuniform across the system.
Monte Carlo simulations have recently shown that the use of such an exchange gradient can be an effective way to design thermodynamic behaviour \cite{salcedo-gallo21}. Moreover, magnetization \cite{borys2020asymmetric,gallardo19b} and thermal \cite{langner18} gradients have also been recently used to modify spin wave dispersion. 
Beyond magnetic systems, heat gradients have been employed to break time-reversal symmetry in phonon systems for angular momentum generation \cite{hamada18}.
Because of this, the concepts developed here will apply broadly to waves in other physical systems, such as mechanical, \cite{fulcher1985study} elastic, \cite{attarzadeh2020experimental} phononic \cite{wang2018observation} and electromagnetic systems, where a similar gradient in physical properties, such as density or index of refraction, could be used to induce unusual phenomena. 

Our idea of using gradient exchange is, in fact, based on a comparison with features found in systems with Dzyaloshinskii–Moriya interactions (DMI){\textemdash}or antisymmetric exchange.  
If one looks, for example, at a one-dimensional system with the magnetization directed along the $z$ axis and propagation along the $x$ axis, then the DMI contribution (with symmetry broken in the $y$ direction) to the $x$ component of the effective field contains a spatial first derivative term such as \cite{moon13}
\begin{equation}
    H_{\textrm{DMI}}^{x} = \frac{D}{\mu_0 M_s}   \frac{\partial {m}^y}{\partial x},    
\end{equation}
 where $D$ is the DMI constant with units of J/m$^2$, $M_s$ is the saturation magnetization, $\mu_0$ is the permeability of free space, and $\mathbf{m}$ is a unit vector in the direction of the local magnetization.
This term ultimately leads to a term in the dispersion relation that scales with wavenumber $k_x$ for small wavenumber (unlike symmetric exchange which contributes a $k_{x}^{2}$ term) and hence leads to the nonreciprocal propagation, where $\omega(+k_x)\neq\omega(-k_x)$.  
In a normal ferromagnet, in contrast, such a linear, first derivative term does not exist.  
However, if the exchange interaction is nonuniform in the $x$ direction, a similar term appears in the  effective field.
For example, the $x$ component of the effective field contains an additional contribution given by
\begin{equation}
    H_{\textrm{new}}^{x} \sim \Delta J \frac{\partial {m}^x}{\partial x},
\end{equation}
where $\Delta J$ represents the spatial change in the exchange constant.  
A nonuniform exchange can be achieved, for example, by doping \cite{sorensen19} or by ion bombardment.
We should note that this term does not produce the chiral groundstates seen in DMI systems but -- as we will see -- it does lead to spatially nonuniform modes, nonreciprocity and asymmetric group velocity.

Nonreciprocity is a key feature of spin waves{\textemdash}and waves in other media \cite{camley83,camley87}. 
This is when the reversal of the direction of propagation leads to different localization of the waves, or even different frequencies \cite{damon61,camley87}.
Nonreciprocity is currently the basis of a variety of technological devices, including diodes, circulators and isolators, that allow unidirectional transmission of signals \cite{lan2015spin,shao20}.

While nonreciprocal propagation in magnets is a well-known phenomenon, it is not present in all magnetic systems.
This can, however, be achieved in many ways, including through engineered dipolar interactions \cite{gallardo19}. 
In the past few years, the interfacial DMI has emerged as a new way to create nonreciprocal spin waves \cite{moon13}.
In ultrathin ferromagnetic films grown on materials with a strong spin-orbit coupling, interfacial DMI exists due to inversion-symmetry breaking \cite{bode2007}. This induces a nonreciprocal spin wave dispersion, \cite{zakeri2010asymmetric,moon13,cortes13} and hence exotic effects such as nonreciprocal propagation, \cite{ma14,garcia2014nonreciprocal,kostylev2014interface} energy focusing \cite{kim2016spin} and the nonexistence of standing waves \cite{zingsem2019unusual,flores2020semianalytical}.

In particular, we explore the following three consequences of gradient exchange on spin waves. 
Firstly, we calculate the eigenmodes of the system with nonuniform exchange. 
These eigenmodes are spatially asymmetric and cannot be represented by a single $k$ wavevector.
In addition, spatial localization of spin waves is found.
Secondly, we investigate the possibility of nonreciprocal excitation of spin waves travelling in opposite directions.
An important issue in the development of nonreciprocal devices is the frequency range of operation. For instance, magnetostatic devices based on yttrium iron garnet are limited to frequencies below 10~GHz for moderate biasing fields, devices based on metallic ferromagnets can operate up to 25~GHz \cite{kuanr2009nonreciprocal}, and hexagonal ferrites have an upper limit typically around 50~GHz \cite{camley2009high,song2009millimeter}.  
In contrast, the method developed in this work produces nonreciprocal behavior at frequencies extending into the hundreds of GHz.
Finally, we show that spin waves propagating in one direction have a different group velocity compared to those propagating in the opposite direction.

\section{Spin wave modes}

\begin{figure}
\includegraphics[width=0.6\linewidth]{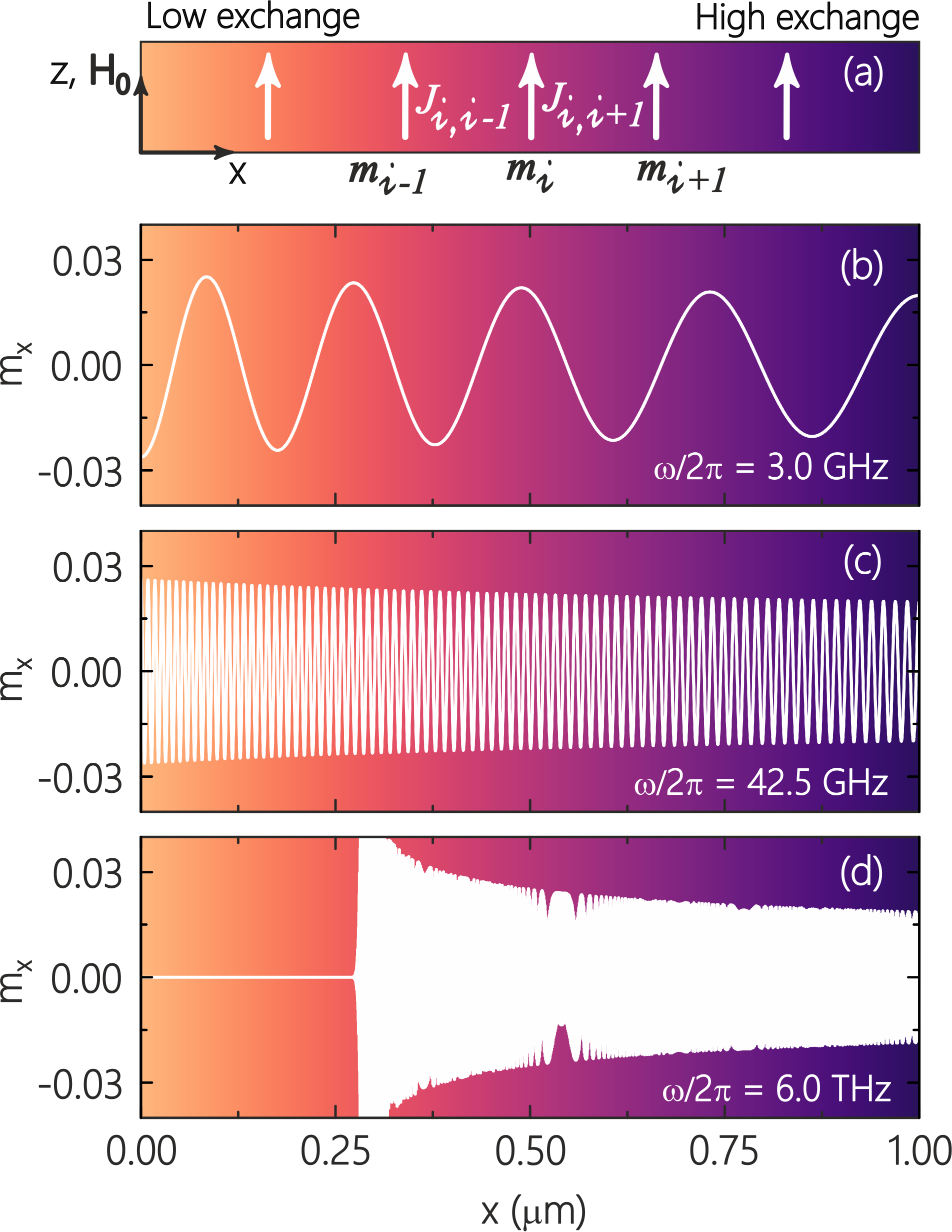}
\caption{
{(a) Schematic of a magnetic strip modelled as a linear chain of exchange-coupled spins indexed by $i$, each with unit vector direction $\textbf{m}_{i}$. A uniform magnetic field $\mathbf{H_0}$ is applied along the $z$ direction. 
The color gradient in the background represents the linearly-varying exchange from low (left) to high (right).
Eigenmodes for $N=$4000 spins (a 1~$\mu$m long strip with nonuniform exchange) at eigenfrequencies $f$ equal to (b) 3.0~GHz, (c) 42.5~GHz, and (d) 6.0~THz.
}}
\label{fig:Modes}
\end{figure}

Before calculating how signals propagate through a system with nonuniform exchange, we find its spin wave modes. 
These can then be used to gain insight into nonreciprocal behavior. 
We consider a one-dimensional magnetic strip; a model system of $N$ exchange-coupled spins  with a uniform magnetic field $\mathbf{H}_0$ applied in the $z$ direction, as shown in Fig.~\ref{fig:Modes}(a). 
The torque equation describing the magnetization dynamics of a spin vector $\mathbf{m}_{i}$ with unit length at site $i$ is given by
\begin{equation}
\label{torque}
    \frac{\partial\mathbf{m}_i}{\partial t} =-|\gamma| \mu_0 \mathbf{m}_i\times\mathbf{H}_{\textrm{eff},i},
\end{equation}
where $\gamma/2\pi=29.2$~GHz/T is the gyromagnetic ratio, and $\mathbf{H}_{\textrm{eff},i}$ is the effective field at site $i$ due to its two neighboring spins and the external applied field. Note that here we have ignored dipolar effects as the magnet is considered small and because at the high frequencies considered, the excitations are exchange-dominated. Damping is also ignored for now. The effective field is given by
\begin{equation}
\label{effField}
    \mathbf{H}_{\textrm{eff},i} = \frac{J_{i,i-1}}{\mu_0} \mathbf{m}_{i-1} + \frac{J_{i,i+1}}{\mu_0} \mathbf{m}_{i+1} + H_{0} \hat{\mathbf{z}},
\end{equation}
where $J_{i,i\pm1}$ denotes the exchange field between a spin at sites $i$ and one at a neighboring site indexed by $i\pm 1$, in units of Tesla. 
We consider a 1~$\mu$m long strip of iron ($N=$4000 sites) with the exchange field $J$ varying linearly from 43.5~T at the left side ($i=1$) to 132~T at the right side ($i=N-1$) and $\mu_0H_0 = 0.1$~T. The largest value of $J$ here is based on that of bulk iron \footnote{The largest value of the exchange field used throughout this letter was estimated from the Curie temperature for Fe of $T_C=$~1043~K, as reported by Ref.\cite{kirby85}}.

Assuming linear, precessional solutions such that $m_{i}^{x/y} \sim e^{-\textrm{i} \omega t}$ and $m_{i}^{z} \sim 1$ means that Eq.~\eqref{torque} results in a set of $2N$ linear, coupled equations. 
These can be solved numerically to find the resonant frequencies $f_{j}$ (eigenvalues, with $j\in [1,N]$) and modes $\vec{A}_{j} = \{ m_{1}^{x}, m_{2}^{x} ,\cdots, m_{N}^{x} \}$ (eigenvectors of length $N$) for the $N$ spins.
With the absence of dipolar effects, there is a two-fold degeneracy for all the modes. 
This process has been applied in the past to study magnets with uniform exchange \cite{moore2014spin} or with exchange and anisotropy that changes abruptly at an interface, \cite{stamps1996spin,krawczyk2001forbidden} such as exchange spring \cite{livesey06}. 
Some representative modes are shown in Fig.~\ref{fig:Modes} at frequencies (b) 3.0~GHz, (c) 42.5~GHz and (d) 6.0~THz. 
We see that the mode shapes are asymmetric and that at very high frequencies, they can even become confined to the right (high exchange) side of the material. Of particular note for what follows is that the wavelength of an eigenmode is shorter on the left versus the right. 
Also note that all the eigenmodes are orthonormal, meaning that they can act as a basis onto which magnetization excitations can be projected in a generalized Fourier decomposition.

\section{Nonreciprocal Excitation}

The observation of asymmetric modes leads one to believe that nonreciprocal propagation may also be observed. 
To test this, we performed Landau-Lifshitz-Gilbert simulations on the atomistic model of $N$ spins described above.
We developed Fortran codes, based on Eq.~\eqref{torque} with the addition of a damping term $-|\gamma|\mu_0 \alpha[\mathbf{m}_i\times(\mathbf{m}_i\times\mathbf{H}_{\textrm{eff},i})]$, with the Gilbert damping paratmeter $\alpha=10^{-4}$, as that of ultralow-damping materials, such as metallic Fe$_x$Co$_{1-x}$ alloys \cite{schoen16}. 
We also added an oscillatory driving field -- spatially localized on one side of the strip or the other (left or right) -- to the effective field in Eq.~\eqref{effField} with a driving frequency $f_d$. 
This field is denoted $\mathbf{h}(x,t) = g(x) \cos(2\pi f_d t) \hat{\mathbf{x}}$, where $g(x)$ is taken to be a square driving profile which turns on at time $t=0$.

Numerical integration was performed using a second-order Runge-Kutta scheme, with timesteps of 10$^{-7}$~ns.
The chain of $N$ spins was driven uniformly along a block of 200 spins ($d=$~50~nm) on either left or right side of the magnet, so $g(x)$ is a square step function in both cases, and the resulting dynamics were recorded.
This resembles a typical spin-wave device comprising antennas placed at either ends of a thin magnetic stripe \cite{obry13,ciubotaru16} generating a driving excitation [see Fig.~\ref{fig:Modes}(a)].  

Typical results illustrating the nonreciprocal behavior are shown in  Fig.~\ref{fig:propagation}(b)-(c) and videos are included in Supplemental Information. 
Two snapshots of the magnetization component $m^x$ as a function of position $x$ are presented after the system is driven for $t=0.7$~ns.
The magnet is driven from the left and the right, as depicted by the white shaded regions on Fig.~\ref{fig:propagation}(b)-(c). 
At the chosen driving frequency $f_R=42.5$~GHz and for this particular driving block width $d$, one sees a strong transmission of signal from the right (c), but not from the left (b). 
This is precisely the behavior that is desired for nonreciprocal devices such as isolators and filters. 
(Interestingly, in panel (b) one sees a higher amplitude at the leading edge of the propagating excitation near $x=0.7$~$\mu$m, although the driving has been applied continuously since $t=0.$)

\begin{figure}
\includegraphics[width=0.6\linewidth]{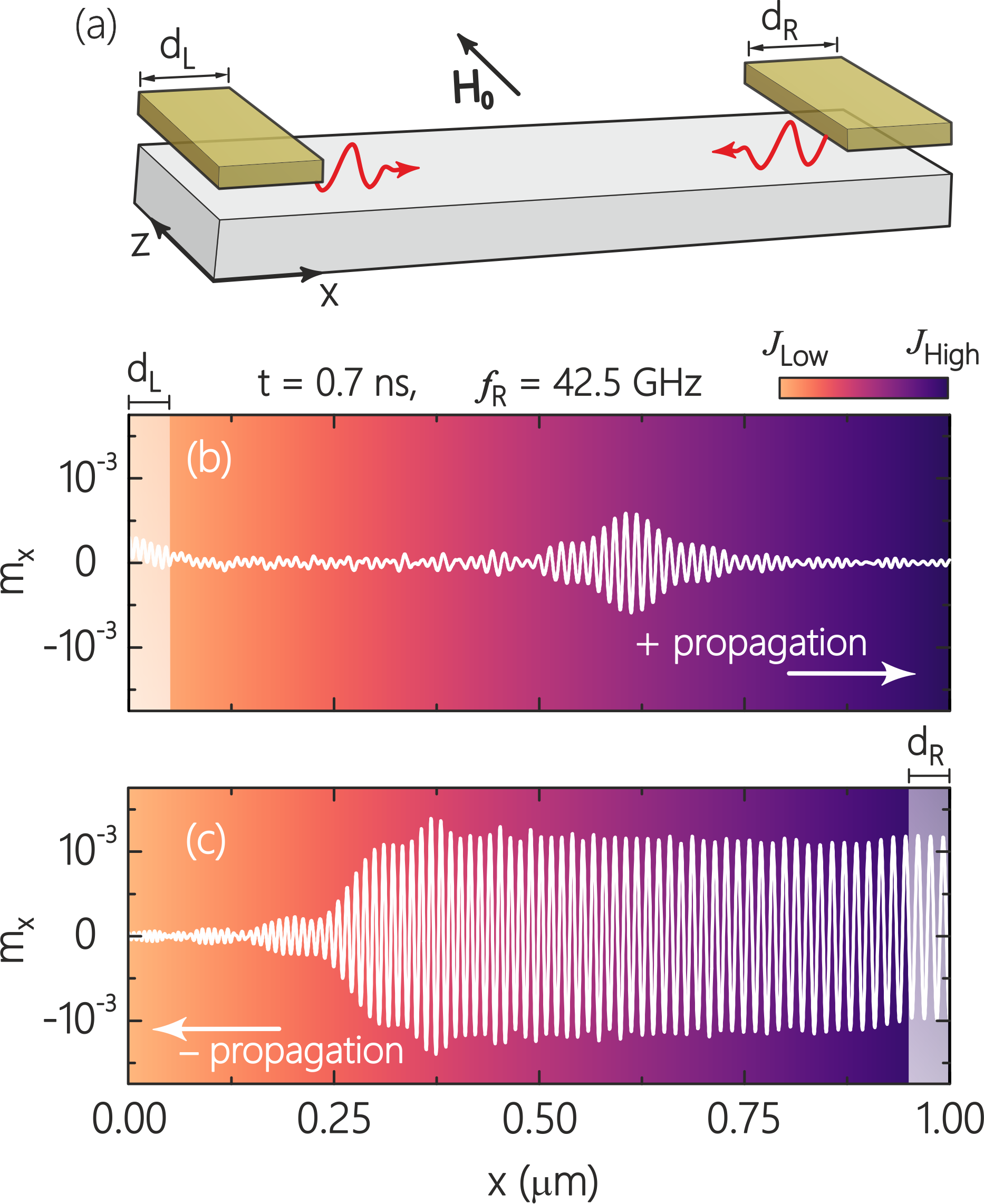}
\caption{
{(a) Schematics of a typical system for spin wave generation in a magnetic strip (gray) with a field applied from an antenna (yellow bar) of width $d$ at either the leftmost end, or the rightmost end of the strip.
Propagation of a signal from the (b) left and (c) right sides of a magnetic strip with varying exchange, at a snapshot in time $t = 0.7$~ns. The driving profile $g(x)$ is a square pulse with width $d=50$~nm corresponding to 200 spins on either the left or right, the driving frequency is $f_d=42.5$~GHz, and the driving field amplitude is 3~mT.
}}
\label{fig:propagation}
\end{figure}

Although the nonreciprocity is clear in this example, and perhaps not surprising given the strong asymmetry in the system exemplified by the modes shown in Fig.~\ref{fig:Modes}(c), its origin is not at first apparent.
To explain it -- and indeed predict at which frequencies it will occur -- we calculate the overlap $\mathscr{O}_j$ between the driving field profile $g(x)$ (given as a discrete vector $\vec{g}$) and each normalized eigenmode $\vec{A}_j$, namely
\begin{equation}
\label{overlap}
    \mathscr{O}_{j} = \vec{g} \cdot \vec{A}_{j} = \sum_{i=1}^{N} g_i A_{j,i},
\end{equation}
where $i$ is the site index and the $i$th component of the driving field profile $g_i$ is taken from either the vector
\begin{equation}
g_{\textrm{left}}(x) \to \vec{g}_{\textrm{left}} = \{ \underbrace{1,1,\cdots,1}_{\textrm{200 terms}} , \underbrace{0,0,0,0,\cdots,0}_{N\textrm{-200 terms}} \},
\end{equation}
or
\begin{equation}
g_{\textrm{right}}(x) \to \vec{g}_{\textrm{right}} = \{   \underbrace{0,0,0,0,\cdots,0}_{N\textrm{-200 terms}},
\underbrace{1,1,\cdots,1}_{\textrm{200 terms}}\}.
\end{equation}
Note that the overlap value is also a coefficient in the discrete, \emph{generalized} Fourier decomposition \cite{generalizedFT} of the driving profile given by $\vec{g} = \sum_{j=1}^{N} \mathscr{O}_{j} \vec{A}_{j}$.
Since each eigenmode has a unique frequency $f_j$, one can plot the overlap value or Fourier coefficient as a function of frequency. 
A large value of $\mathscr{O}_j$ indicates strong coupling to a particular driving field profile. 
Driving at the eigen-frequency corresponding to an eigenmode with strong coupling will then result in efficient signal generation and propagation, while choosing frequencies corresponding to weak coupling results in little excitation in the magnet.

\begin{figure}
\includegraphics[width=0.6\linewidth]{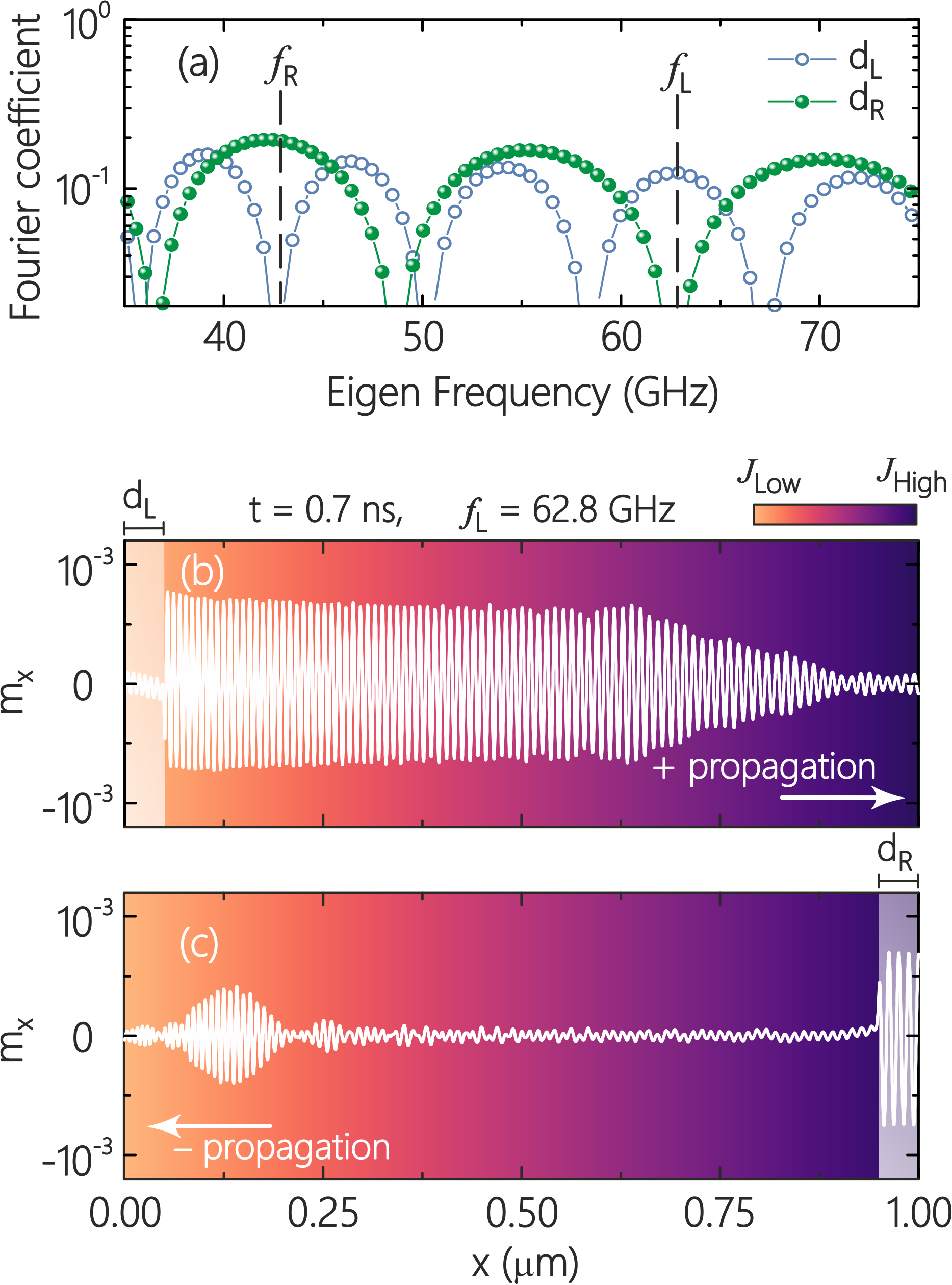}
\caption{(a) Generalized Fourier coefficients (overlap values $\mathscr{O}_j$) as a function of mode frequency for right- (green, solid circles) and left- (blue, open circles) driven systems with varying exchange.
The frequencies marked are $f_R=42.5$~GHz and $f_L=62.8$~GHz.
The $m^x$ component along the strip is shown at a snapshot in time $t=0.7$~ns for signal driven from the (b) left and (c) right. The strip is driven with a square pulse with width $d=50$~nm corresponding to 200 spins, at frequency $f_L$, and the driving field amplitude is 3~mT.
}
\label{fig:Decomp_omegaL}
\end{figure}

Now the significance of the eigenmode asymmetry becomes apparent: driving with a pulse that is $d$ wide may couple strongly with an eigenmode when applied to the left, but does not necessarily couple strongly with that same eigenmode when applied to the right. This is because the eigenmode has a very different wavelength -- and net moment across a distance $d$ -- on the left compared to on the right.

In Fig.~\ref{fig:Decomp_omegaL}(a) we plot the generalized Fourier coefficients $\mathscr{O}_{j}$ for right- (green, solid balls) and left- (blue, open balls) driven systems, versus the eigenfrequencies. 
The magnetic parameters are all the same as used in Fig.~\ref{fig:propagation}, and once again $d=50$~nm. 
One sees that the overlap values oscillate in size. Large amplitudes mean an efficient excitation at that frequency. 
In particular, we have marked two frequencies of interest at $f_R = 42.5$~GHz (used to make Fig.~\ref{fig:propagation}) and $f_L = 62.8$~GHz. 
We predict using the argument just detailed that driving the system with frequency $f_R$ should result in efficient excitation on the right, but not on the left, due to its large overlap with the driving profile $\vec{g}_{\textrm{right}}$ and small overlap with $\vec{g}_{\textrm{left}}$, as demonstrated by the size of the $\mathscr{O}_j$ values. This is indeed what we found when performing the numerical LLG experiment, as was discussed in Fig.~\ref{fig:propagation}(b)-(c).

Similarly, Fig.~\ref{fig:Decomp_omegaL}(a) explains that there should be a far more efficient excitation at $f_L=62.8$~GHz when the driving occurs on the left (blue, open balls) rather than on the right (green, solid balls). 
This again is supported through our numerical experiments, as shown in Fig.~\ref{fig:Decomp_omegaL}(b)-(c). Two snapshots of the magnetization component $m^x$ at time $t=0.7$~ns are plotted along the magnetic strip's length. The end of the strip has been driven at $f_L$ since time $t=0$ from the left (b) and from the right (c), over regions that are 50~nm long. This time, as predicted by the overlap values $\mathscr{O}_j$, driving from the left produces a larger propagating excitation. Hence, by changing the frequency one can find intermittent windows where signals can be excited from the left but not the right, and vice-versa.

\section{Nonreciprociprocity at high frequencies}

\begin{figure}
\includegraphics[width=0.6\linewidth]{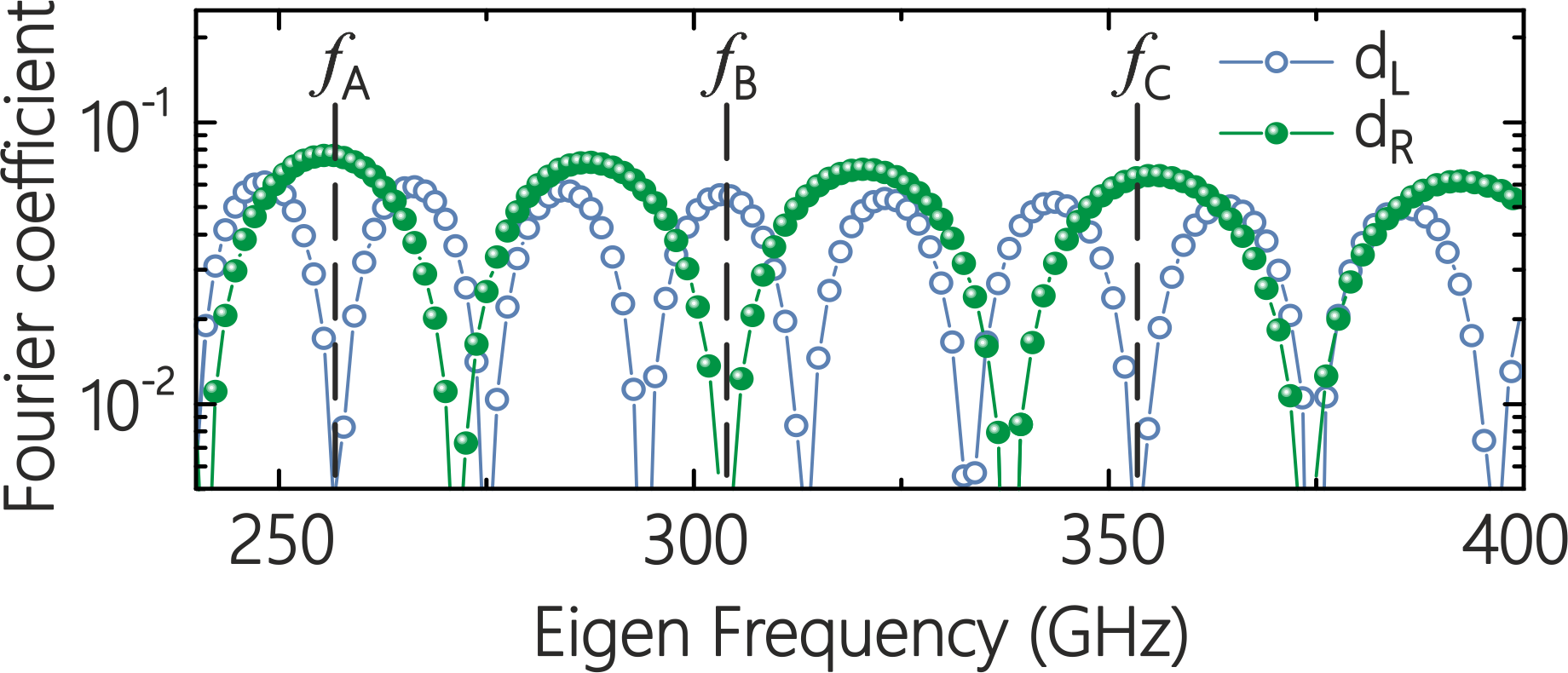}
\caption{$\mathscr{O}_j$ for right- (green, solid circles) and left- (blue, open circles) driven systems with nonuniform exchange at high frequencies. The parameters used to calculate these are the same as those in Fig.~\ref{fig:Decomp_omegaL}.
Lines mark frequencies $f_A=257$, $f_B=304$, and $f_C=353$ GHz, where nonreciprocal excitation should take place.
}
\label{fig:HighFreqDecomp}
\end{figure}

What is particularly exciting is the fact that the nonreciprocal frequencies can be tuned to occur anywhere across the spin wave Brillouin zone, with frequencies ranging from the low GHz to the THz. 
As Figs.~\ref{fig:propagation} and \ref{fig:Decomp_omegaL} showed, nonreciprocity occurs, for example, around 40-60 GHz, and without the need to apply a large magnetic field. Typically, such frequencies are too high for ferromagnetic materials with small applied fields. 
However, such frequencies are important because they lie in a band which is becoming important as the low GHz microwave bands become increasingly full \cite{dhillon20172017}. 
Nonreciprocal driving is also seen at much higher frequencies, in the hundreds of GHz.
In Fig.~\ref{fig:HighFreqDecomp} we show the generalized Fourier coefficients $\mathscr{O}_j$ for a system with nonuniform exchange at frequencies between 240 and 400 GHz, using the same material parameters as so far have been used throughout this work.
The vertical lines at frequencies $f_A=257$, $f_B=304$, and $f_C=353$ GHz indicate regions where maximum nonreciprocity should be observed, where efficient driving occurs only at the right, left, and right respectively. Notice that the $\mathscr{O}_j$ are plotted on a logarithmic scale so the efficiency of exciting from the right or the left is predicted to be very different at these points.

Note that the driving field profile $g(x)$ is assumed uniform across a region of length $d$ in the situations described above, and is zero elsewhere. 
Such a step function for the driving field is not realistic in an experiment. 
However, the generalized Fourier decomposition into spin wave eigenmodes is completely general and a more realistic profile $g(x)$ can be inserted into Eq.~\eqref{overlap} to find the special frequencies at which a signal will propagate from one end and not the other.
The calculation of the coefficients or overlap values $\mathscr{O}_j$ is extremely fast as two $N$-dimensional vectors are multiplied together.
We do note, however, that the calculation of the eigenmodes $A_j$ can be computationally demanding as $N$ becomes increasingly large. 
We also point out that this calculation must be done atomistically in order to recover correct results at high frequencies.

\section{Implications on Group Velocity}

Having examined how non-uniform exchange leads to nonreciprocal behavior---a phenomena which, as mentioned in the introduction, is also a distinct feature of DMI systems---we now turn to another analogy between these two systems: the group velocity of spin waves.
The group velocity is defined as
\begin{equation}
    \vec{v}_g = \frac{d\omega}{d \vec{k}}.
    \label{eq:vg}
\end{equation}
An intriguing aspect of interfacial DMI is that it can generate asymmetric $v_g$ \cite{moon13}. For instance, recent experimental work by Wang and co-worker \cite{wang20} demonstrated that spin waves in ultrathin YIG films propagating in opposite directions but with the same wavenumber ($+k$ and $-k$) have different group velocities.

This is a direct consequence of the term linear in $k$ (for small wavenumbers) appearing in the spin wave dispersion of DMI systems. Our system produces an analagous effect. In the small $k$, exchange-dominated limit, one can relate $v_g$ to frequency using $\omega = \mathscr{D} k^2$ and Equation~\eqref{eq:vg} which gives
\begin{equation}
    v_g \approx 2\sqrt{\omega \mathscr{D}},
    \label{eq:vgNew}
\end{equation}
with $\mathscr{D}$ a modified version of the exchange constant. From this, we see that a large $\mathscr{D}$ leads to large $v_g$, whereas small exchange would induce slower wave propagation. In a material with a gradient in the exchange, this means that spin waves moving left from an excitation point will have a different group velocity than those moving right from that same point.

\begin{figure}[tb]
\centering
\includegraphics[width=0.6\linewidth]{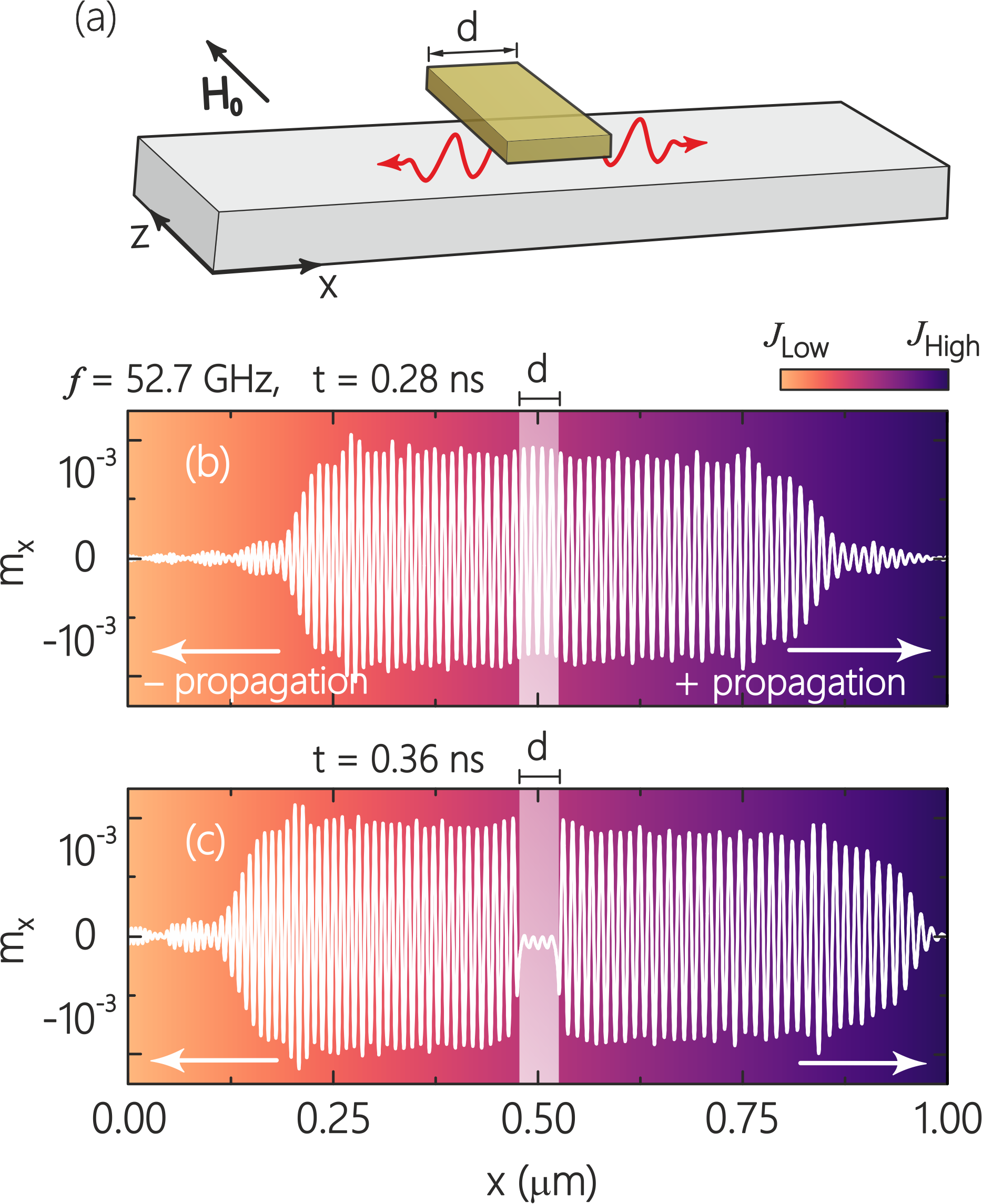}
\caption{(a) Schematics of a typical system for spin wave generation in a magnetic strip (gray) with a field applied from an antenna (yellow bar) of width $d$ placed at the centre of the strip.
Propagation of a signal from the centre of a magnetic strip with varying exchange, at a snapshot in times (b) $t = 0.28$~ns and (c) $t = 0.36$~ns. The driving profile $g(x)$ is a square pulse with width $d=50$~nm corresponding to 200 spins, the driving frequency is $f_d=42.5$~GHz, and the driving field amplitude is 3~mT.}
\label{fig:Centre}
\end{figure}

In order to test this, we have revised our numerical experiments discussed in relation to  Figs.~\ref{fig:propagation}-\ref{fig:Decomp_omegaL}. 
The geometry is changed slightly so that the exciting region with width $d$ is now at the center of the chain of dipoles, as depicted in Fig.~\ref{fig:Centre}(a). Numerical integration results are shown in Figs.~\ref{fig:Centre}(b) and \ref{fig:Centre}(c) at 0.28~ns and 0.36~ns, respectively.
Much like the previous figures, the chain is 4000 spins long (approximatly 1~$\mu$m) and is driven uniformly along a block of 200 spins ($d=$~50~nm) at its center.
At 0.28~ns, one can already see that the excitation moving rightward is closer to the edge than the wave going leftward, this becomes even more evident at 0.36~ns when the wave going right has reached the edge and the wave moving left has not.
This is in agreement with our predictions from Eq.~\eqref{eq:vgNew} that high exchange regions (right) will generate faster propagating waves than lower exchange regions (left).
To quantify this, we take the example shown in Fig.~\ref{fig:Centre}(c) and calculate the velocity of the excitation propagating in both directions. 
We find that $v^{-} \approx 1000$~m/s and $v^{+}\approx 1250$~m/s, yielding $\delta v = v^{+} - v^{-} \approx 250$~m/s. This is somewhat higher than the reported drift group velocity in YIG system of $\delta v_g\approx 40.8$~m/s induced by the DMI interaction \cite{wang20}.

It is then tempting to speculate that because of the control of the group velocity, a structure with a nonuniform exchange could be used to structure spin wave pulses.
We must note that although there is an asymmetric propagation from the center, this is not true ``nonreciprocal propagation" as seen in DMI systems. In DMI systems, there is a different velocity moving left versus right at \emph{the same position in space}. 
In the example just described here and illustrated by Fig.~\ref{fig:Centre}, this is not the case and it is only because the left- and right-moving excitations enter different spatial regions with different values of the exchange that their propagation speeds are different. 
In the same sense, the quoted velocities above are simple averages over a time span and these will change with the distance travelled, i.e. change in the relative exchange gradient.


\section{Conclusion}

This article details a way to introduce space inversion symmetry-breaking in magnetic systems through nonuniform exchange interaction. 
Using material parameters based on iron, we show that the spin wave modes in such a system are highly asymmetric and can become localized at large frequencies in the low THz range.
Our theoretical approach provides evidence of nonreciprocal excitation of spin waves as well as predicts at what frequencies this nonreciprocity occurs. 
Of particular note is the way that driving fields can be chosen to generate nonreciprocal excitation at desired frequencies ranging from low to high GHz. 

Furthermore, the similarities and differences in excitation propagation between nonuniform exchange systems and systems with DMI are discussed{\textemdash}by examining the group velocity of spin waves we have found that left and right propagation from a region are markedly different.
These observations have implications on magnonic and signal processing applications where spin wave propagation must be controlled. Our findings also have consequences in the field of `spin caloritronics' which is concerned with heat propagation across magnetic systems for applications such as unidirectional spin wave heat conveyers \cite{an13}.

\begin{acknowledgments}
R. Mac\^edo acknowledges support from the Leverhulme Trust and the University of Glasgow through LKAS funds as well as the hospitality of The University of Colorado at Colorado Springs where part of this work was performed.
\end{acknowledgments}

\nocite{*}

\providecommand{\noopsort}[1]{}\providecommand{\singleletter}[1]{#1}%

\end{document}